\documentclass[12pt]{article}
\usepackage{graphicx}

\begin{document}

\title{Stochastic Penna model for biological aging}
\author{Zhi-Feng Huang\thanks{E-mail address: 
zfh@thp.uni-koeln.de; Address after 1 Nov. 2000: 
Department of Physics, University of Toronto,
Toronto ON M5S 1A7, Canada}
and Dietrich Stauffer\thanks{E-mail address: 
stauffer@thp.uni-koeln.de}\\
\small{Institute for Theoretical Physics, Cologne University, 
50923 K\"{o}ln, Germany}}
\date{}
\maketitle

\begin{abstract}
A stochastic genetic model for biological aging is introduced
bridging the gap between the bit-string Penna model and the
Pletcher-Neuhauser approach. The phenomenon of
exponentially increasing mortality function at intermediate ages
and its deceleration at advanced ages is reproduced for
both the evolutionary steady-state population and the genetically
homogeneous individuals.
\vskip 12pt
{\em Keywords}: Biological Aging; Mortality; Penna Model.
\end{abstract}

\section{Introduction}
\label{sec1}

The problem of biological aging has attracted much attention in
recent years. Based on the data of human demography and experiments
of other living organisms, many important phenomena of longevity 
have been found \cite{book,vaupel98,pletcher}. For instance, the
Gompertz law was observed for intermediate ages, that is,
the mortality function increases exponentially with age, while 
at old ages the mortality was found to decelerate or level off, 
and even decline for some organisms like flies, worms, and yeast 
\cite{vaupel98,curtsinger92, perls98}.

To reproduce and explain these phenomena, various models of
senescence have been proposed, with genetic or nongenetic mechanisms 
\cite{book,vaupel98,pletcher,vaupel93,penna}. Among them, the
one widely used by physicists is the Penna model \cite{penna,book},
where one computer word is used to represent the inherited genome 
of one individual and each bit of the word corresponds to one age 
of the individual lifetime. A bit set to one represents a
deleterious mutation and the suffering from an inherited disease
from this age on, and the individual will die if the accumulation
of these set bits exceeds a threshold.

Although the Penna model has been well applied to many problems
related to biological aging \cite{book,p-m95}, there exists an 
important flaw in this model as pointed out by Pletcher and
Neuhauser very recently \cite{pletcher}. That is, the model 
predicts that for a genetically identical population all 
individuals have their genetic death at the same age, but this
is inconsistent with the experimental results \cite{curtsinger92,
vaupel98} which also exhibited the exponential Gompertz law
and the deceleration of the old age mortality for the genetically
homogeneous case. Thus, a more complicated model has been proposed
\cite{pletcher}.

In this paper we develop a simpler stochastic model bridging the
gap between the standard (deterministic) Penna model
and the Pletcher-Neuhauser approach. The simulations and 
analytic results of this model are shown to agree with some 
features of the biological aging, e.g., the exponential increase
of the mortality function and the deceleration at advanced ages, and 
the flaw of the Penna model mentioned above can be avoided.

\section{Model}
\label{sec2}

As in the standard Penna model, here the genome of each individual
is characterized by a string (computer word) of 32 bits, and each
bit is expressed as a particular age in the life of the individual. 
A bit $i$ is set to $1$ if it represents a deleterious mutation,
and from this age $i$ on this bit will continuously affect the 
survival probability of the individual. That is, at age $a$
($\geq i$) the death probability contributed by the mutated bit $i$
is $f(a-i)$, with the corresponding survival probability $1-f(a-i)$.
Otherwise, this bit is set to zero and has no effect on death.
Thus our assumptions are very different from an earlier "Fermi"
function in another stochastic Penna model \cite{fermi}.

The individual's survival probability $G$ up to age $a$ is the
product of the contributions from all the bits before $a$:
\begin{equation}
G(1,2,...,a)=(1-b_1 f(a-1))(1-b_2 f(a-2))\cdots(1-b_i f(a-i))
\cdots(1-b_a f(0)),
\label{liveprob}
\end{equation}
where $b_i =1$ or $0$ ($i=1,...,a$) represents the $i$th bit.
With the form of $f(a-i)$, one can obtain the mortality function
by simulation or analytical work. In this work we simply assume 
that
\begin{equation}
f(a-i)=(a-i+1)C,
\label{deathfunc}
\end{equation}
with the constant $C=0.03$ and the limit $f\leq 1.0$, which means 
that the contribution of death probability from bit $i$ (if set to 
$1$) is assumed to increase linearly with the age. The other forms
of $f(a-i)$, such as the exponential and the square root forms, have
been tried, and we have also simulated the other probabilistic 
Penna model with Fermi function \cite{fermi}. Although some
phenomena for the genetically heterogeneous steady-state population
can be reproduced, they cannot give a good result for the
genetically homogeneous populations.

The alive individual will generate $B$ offsprings from the minimum 
reproduction age $R_{\rm min}$ to the maximum one $R_{\rm max}$,
and the genome of each offspring is the same as the parent one,
except for $M$ mutations randomly occurring at birth. At each time
step $t$, a Verhulst factor $V=1-N(t)/N_{\rm max}$ denoting the 
survival probability of the individual due to the space and food 
restrictions is introduced, where $N(t)$ is the current population
size and $N_{\rm max}$ is the carrying capacity of the environment,
usually set to $10 N(0)$.
In the next section \ref{sec3} the simulations based on these
rules are presented, while for genetically identical individuals,
which have the same genotype randomly sampled from the simulated 
steady-state population, the analytic results can be derived, as 
shown in section \ref{sec4}.

Moreover, in this paper the mortality function $\mu (a)$ at age 
$a$ is defined as
$$\mu (a) = -{\frac {d\ln N_a}{da}}\simeq -\ln S(a),$$
where $N_a$ denotes the number of alive individuals with age $a$,
and $S(a)=N_{a+1}/N_a$ is the survival rate. To eliminate the
effect of the Verhulst factor, the normalized mortality function
is preferred \cite{book}, i.e.,
\begin{equation}
\mu (a) = -\ln [S(a)/S(0)].
\label{mortfunc}
\end{equation}

\section{Simulations}
\label{sec3}

In our simulations, initially the population size $N(0)$ is 
$10^7$ and all bits of all the strings are set to zero, i.e.,
free of mutations. One time step $t$ corresponds to one aging
interval of the individuals, or reading one bit of all strings. 
The reproduction range is set from $R_{\rm min}=6$ to 
$R_{\rm max}=20$ with the birth rate $B=1$, and the results
are similar if using the maximum value of $R_{\rm max}=32$. 
$M=1$ mutation for each offspring genome is introduced at birth, 
and here only the bad mutations are taken into account, that is, 
the bit randomly selected for mutation is always set to $1$.
(The good mutations have also been considered, e.g., 10\% good
mutations and 90\% bad ones, and similar results are found.)

\begin{figure}
\centerline{
\resizebox{0.8\textwidth}{!}{%
  \includegraphics[angle=-90]{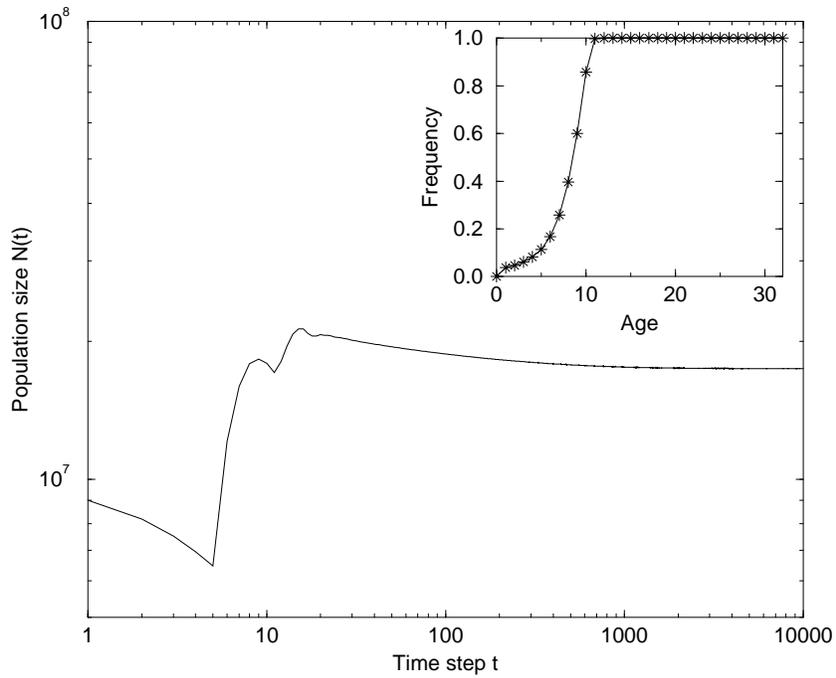}}}
\caption{The evolution of the whole population size $N(t)$
with time step $t$, for the initial population $N(0)=10^7$ and
the parameters $R_{\rm min}=6$, $R_{\rm max}=20$, $B=1$, and
$M=1$. Only the bad mutations are considered. Inset: 
the frequency of deleterious bits (set as $1$) as a function of
age for the individuals of the steady-state population (averaged 
over timestep 5000 to 10000).}
\label{fig-pop-mut}
\end{figure}

Fig. \ref{fig-pop-mut} shows the evolution of the whole 
population size $N(t)$ until $t=10^4$. Similar 
to the standard Penna model, the steady-state 
population is obtained at late timesteps, and as a result of 
evolution and selection, the frequency of deleterious bits 
(set as $1$) for the individual of the steady-state population 
is low at early ages (especially before the reproduction age) 
and very high at old ones. This behavior of the frequency 
(or the bad mutation rate) is shown in the inset of Fig. 
\ref{fig-pop-mut}.

\begin{figure}
\centerline{
\resizebox{0.7\textwidth}{!}{%
  \includegraphics[angle=-90]{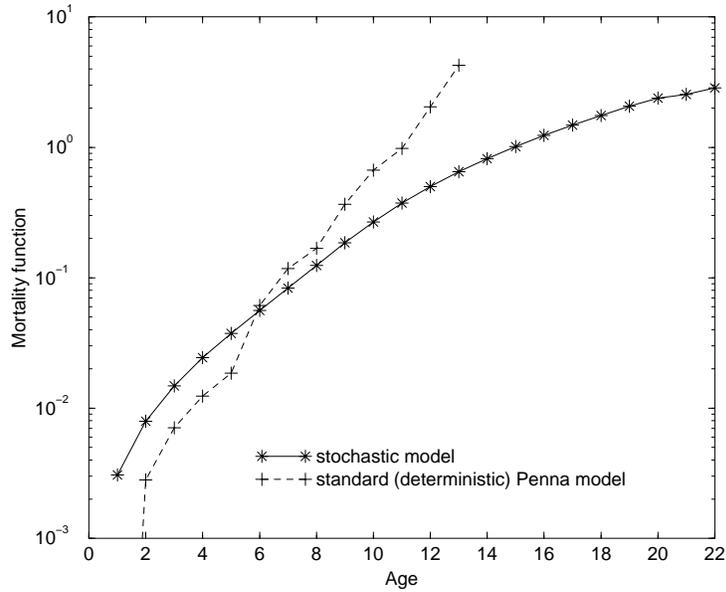}}}
\caption{Linear-log plot of the mortality function for the 
evolutionary steady-state population, with the same parameters 
of Fig. \ref{fig-pop-mut} and averaged over timestep 5000 to 10000. 
The mortality of the standard (deterministic) Penna model is also 
shown (pluses) for comparison, with the same parameters as well as
the death threshold $T=3$.}
\label{fig-mortality}
\end{figure}

The mortality function is calculated using Eq. (\ref{mortfunc})
and averaged over the steady-state population from timesteps
5000 to 10000, as shown in Fig. \ref{fig-mortality}. The result
is consistent with the experimental and empirical observations
\cite{book,vaupel98},
that is, at intermediate ages the mortality function 
increases exponentially, exhibiting the Gompertz law, and
deceleration occurs for old ages. For comparison, the
mortality simulated by the standard (deterministic) Penna model 
is also shown in Fig. \ref{fig-mortality}, with the threshold of the 
accumulated bad mutations $T=3$ and the other parameters unchanged.
The exponential Gompertz law can also be obtained for the
standard Penna model \cite{book}, however, no deceleration is
observed except for suitable modifications summarized in \cite{book};
see also \cite{fermi}.

\section{Genetically identical population}
\label{sec4}

To study the genetically homogeneous population, one can randomly
sample an individual (genotype) from the simulated steady-state
population, and then "clone" it to create the whole genetically
identical population. According to the form of
these bit-strings, the mortality function can be derived and
calculated analytically.

As in some experiments of fruit flies \cite{curtsinger92}, 
reproduction is prevented during the aging of genetically 
homogeneous individuals. Thus, for this population of single 
genotype, we have
\begin{eqnarray}
&N_1&=N_0 (1-b_1 f(0))=N_0 G(0), \nonumber\\
&N_2&=N_1 (1-b_1 f(1))(1-b_2 f(0))=N_1 G(1,2), \nonumber\\
& &\vdots \nonumber\\
&N_a&=N_{a-1} (1-b_1 f(a-1))(1-b_2 f(a-2))\cdots(1-b_a f(0))
     =N_{a-1} G(1,2,...,a), \nonumber\\
& &\vdots \nonumber
\end{eqnarray}
where $N_a$, $a=1,2,...,32$, is the number of individuals with
age $a$ in the population, and the function $G(1,2,...,a)$
is defined by Eq. (\ref{liveprob}). Then the survival rate
is easily obtained:
\begin{equation}
S(a)=\frac {N_{a+1}}{N_a}=G(1,2,...,a+1).
\end{equation}
For the mortality function, the normalized formula
(\ref{mortfunc}) is used to be consistent with the
simulations in Sec. \ref{sec3}, and then we have
\begin{equation}
\mu (a)=-\ln [G(1,2,...,a+1)/G(1)].
\label{hommort}
\end{equation}

\begin{figure}
\centerline{
\resizebox{0.7\textwidth}{!}{%
  \includegraphics[angle=-90]{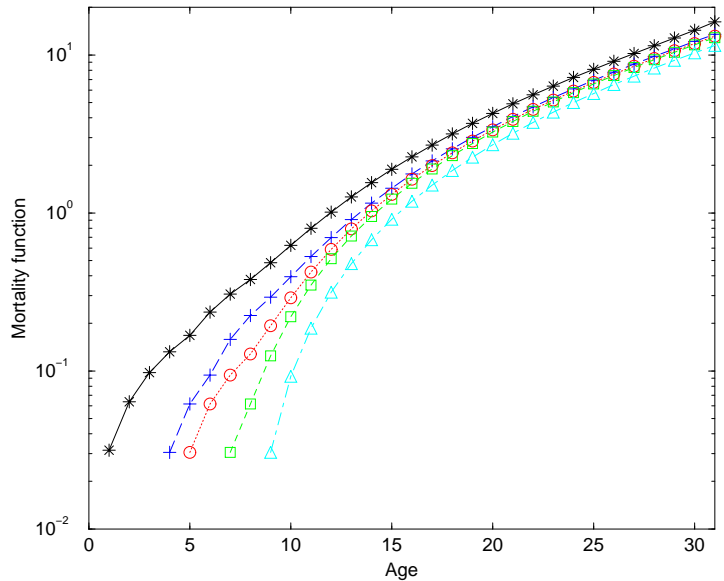}}}
\caption{The mortality function for the genetically identical
populations, with each genotype randomly sampled from the
steady-state population of the simulation shown in Figs. 
\ref{fig-pop-mut} and \ref{fig-mortality}. The results are
calculated by Eq. (\ref{hommort}), and shown in the linear-log
plots.}
\label{fig-clonemort}
\end{figure}

Different genotypes have been selected randomly from the
stable population of Sec. \ref{sec3}, and the corresponding
mortality function of each type is calculated using Eq.
(\ref{hommort}). Some examples are shown in Fig. 
\ref{fig-clonemort} for linear-log plots, where part of them
obey the exponential Gompertz law at the intermediate ages,
similar to that of the above simulation (Sec. \ref{sec3})
and experiments \cite{vaupel98,curtsinger92}. Moreover,
all of these curves exhibit the deceleration for old ages.

Moreover, the analytic calculation is also available if
the reproduction is allowed as in other experiments of 
genetically identical population, but for the case of no 
mutation. The details are shown in the appendix, and the 
mortality function derived is the same as Eq. (\ref{hommort}).

\section{Discussion and conclusion}
\label{sec5}

In this paper a stochastic genetic model of aging is developed
based on the bit-string asexual Penna model, and the results of 
the exponentially increasing mortality at intermediate ages and
its deceleration at old ages are obtained for both the
genetically heterogeneous steady-state population and the 
homogeneous individuals. However, the decrease of mortality
for the oldest ages, observed in some experiments \cite{vaupel98}, 
cannot be described by the mechanism of this model.

Although the properties for intermediate and old ages
have been well simulated in this model, the behavior at early 
ages cannot be well reproduced, which is also an artifact of 
the Penna-type genetic models. From Fig. \ref{fig-clonemort} for
genetically identical populations, it can be found that some
populations have unrealistic zero mortality at some early ages.
Thus, the effects for the early ages studied in the experiments,
such as the investigations of genetic variation for ln-mortality
contributed by steady-state population or by new mutations
\cite{pletcherHC}, cannot be produced in this model.
More efforts should be made to avoid this difficulty, e.g.,
by considering different kinds of genes before and after the 
reproduction age \cite{nkc}.

\section*{Acknowledgements}

We thank Scott D. Pletcher and
Naeem Jan for very helpful discussions and comments.
This work was supported by SFB 341.

\section*{Appendix}

Here an example of the analytic solution for this stochastic model
is presented, for the case where the reproduction is allowed
in the aging process of genetically identical population, but
no mutation occurs when generating the genomes of offsprings.
Thus, the individuals keep homogeneous, characterized by
the same bit-string $b_1 b_2 ... b_L$ with $L$ the length of
genome ($L=32$ in above studies).

When the system evolves to the steady state,
the population size at timestep $t$ of this state
\begin{equation}
N(t)=N_0(t)+N_1(t)+\cdots +N_L(t)
\label{A1}
\end{equation}
as well as the Verhulst factor $V$ can be considered as constant.
Thus, the numbers of individuals with ages from $1$ to $L$ at this 
step $t$ are
\begin{eqnarray}
&&N_L (t)=N_{L-1}(t-1)VG(1,2,...,L), \nonumber\\
&&N_{L-1}(t)=N_{L-2}(t-1)VG(1,2,...,L-1), \nonumber\\
&&\mbox{}\hskip 10mm \vdots \nonumber\\
&&N_a (t)=N_{a-1}(t-1)VG(1,2,...,a), \nonumber\\
&&\mbox{}\hskip 10mm \vdots \nonumber\\
&&N_1 (t)=N_0(t-1)VG(1),
\label{A2}
\end{eqnarray}
where $G(1,2,...,a)$ is the living probability of individual at 
age $a$, as defined in Eq. (\ref{liveprob}), and the individuals
of age zero (newly born) are generated by the ones with 
reproducible age (from age $R_{\rm min}$ to $R_{\rm max}$),
that is,
\begin{eqnarray}
N_0(t)&=&B[N_{R_{\rm min}}+N_{R_{\rm min}+1}+\cdots
+N_{R_{\rm max}}] \nonumber\\
&=&BV[N_{R_{\rm min}-1}(t-1)G(1,2,...,R_{\rm min})
+N_{R_{\rm min}}(t-1)G(1,2,...,R_{\rm min}+1) \nonumber\\
&+&\cdots+N_{R_{\rm max}-1}(t-1)G(1,2,...,R_{\rm max})]
\label{A3}
\end{eqnarray}
with the birth rate $B$.

Consequently, the number of individuals with certain age $a$
($0<a\leq L$) can be expressed as
\begin{equation}
N_a(t)=N_0(t-a) V^a G(1)G(1,2)\cdots G(1,2,...,a).
\label{A4}
\end{equation}
Therefore, if $N_0(t)$ is unchanged for the steady state, 
all the $N_a(t)$, $a=1,...,L$, will also keep unchanged, i.e., 
independent of timestep $t$, and then the survival rate $S$
can be obtained from Eq. (\ref{A2}), that is,
$$S(a)=N_{a+1}/N_a=VG(1,2,...,a+1)$$
and
$$S(0)=N_1/N_0=VG(1).$$
The Verhulst factor can be eliminated when calculating the
normalized rate:
$$S(a)/S(0)=G(1,2,...a+1)/G(1),$$
and from the definition of Eq. (\ref{mortfunc}) one can obtain
the normalized mortality function, which is the same as Eq.
(\ref{hommort}).

The constant property of the population size $N(t)$ and the
number $N_0(t)$ for age $0$, as well as the above analytic result
of the mortality function have been confirmed by the simulation.
Moreover, the steady state condition can be derived from
Eqs. (\ref{A3}) and (\ref{A4}), which is
\begin{eqnarray}
&&BV^{R_{\rm min}}G(1)G(1,2)\cdots G(1,2,...,R_{\rm min})
[1+VG(1,2,...,R_{\rm min}+1) \nonumber\\
&&+V^2G(1,2,...,R_{\rm min}+1)
G(1,2,...,R_{\rm min}+2)+\cdots \nonumber\\
&&+V^{R_{\rm max}-R_{\rm min}}
G(1,2,...,R_{\rm min}+1)\cdots G(1,2,...,R_{\rm max})]=1,
\label{A5}
\end{eqnarray}
depending on the parameters $B$, $R_{\rm min}$, and $R_{\rm max}$.

\newpage

\end{document}